\documentclass[aps,pre,reprint,
  superscriptaddress,
  floatfix,
  nofootinbib
]{revtex4-2}

\usepackage{amsmath,amssymb,bm}
\usepackage{graphicx}
\usepackage[hidelinks]{hyperref}
\usepackage{booktabs}
\usepackage{color}

\newcommand{\tr}{\operatorname{tr}}
\newcommand{\dev}{\operatorname{dev}}

\begin{document}

\title{Polymer viscoelasticity from an orientational closure with implicit stretch}

\author{Souta Miyamoto}
\email{s.miyamoto@cheme.kyoto-u.ac.jp}
\affiliation{Department of Chemical Science and Engineering, Kyoto University,
Kyoto 615-8510, Japan}

\author{Takeshi Sato}
\affiliation{Advanced Manufacturing Technology Institute, Kanazawa University, Kanazawa 920-1192, Japan}

\author{Wanxiang Jin}
\affiliation{Department of Chemical Science and Engineering, Kyoto University,
Kyoto 615-8510, Japan}

\author{Katsuaki Tanabe}
\affiliation{Department of Chemical Science and Engineering, Kyoto University,
Kyoto 615-8510, Japan}
\affiliation{Kyoto MPI Inc., Kyoto 606-8501, Japan}

\author{John J. Molina}
\affiliation{Department of Chemical Science and Engineering, Kyoto University,
Kyoto 615-8510, Japan}

\author{Takashi Taniguchi}
\affiliation{Department of Chemical Science and Engineering, Kyoto University,
Kyoto 615-8510, Japan}

\date{\today}

\begin{abstract}
We propose a deviatoric-stress constitutive equation derived from orientational kinetics, while implicitly retaining contributions from conformational stretch.
Starting from the dumbbell model, we show that conformational stretch affects orientational dynamics through an effective stress scale and rotational diffusivity, which determine an effective relaxation time.
The orientation-based equation is derived by assuming an approximate relation between orientational anisotropy and the stretch state.
Within the proposed framework, constant effective coefficients yield steady--oscillatory shear correspondences closely related to the empirical Cox--Merz and Gleissle--Osaki relations.
For model-specific effective coefficients, the upper-convected Maxwell model is characterized by a constant relaxation time and an increasing modulus under steady shear, whereas finite extensibility with Peterlin preaveraging shortens the relaxation time and suppresses the modulus increase.
With coefficients determined from steady-shear states, the proposed equation successfully reproduces stress responses under transient shear, large-amplitude oscillatory shear, and uniaxial extension.
These results show that the effects of conformational stretch on orientational kinetics can be retained implicitly, providing a constitutive description based on a deviatoric-stress state reconstructable from standard rheometry and a physical interpretation of how stretch enters polymer viscoelasticity.
\end{abstract}

\maketitle
\section{Introduction}

Constitutive equations relate stress to deformation history and are used to predict viscoelastic fluid flows. 
Such constitutive models have been developed from phenomenological considerations~\cite{Oldroyd1958-db} and from microscopic or mesoscopic molecular theories~\cite{Bird1987-la,Bird1995-rx,Doi1988-cc,Larson1988-nn,Doi2013-lv}. 
More recently, data-driven approaches have provided systematic ways for constructing constitutive descriptions from rheometric measurements or simulation data~\cite{Brunton2019-jf,Lennon2023-xc,Shanbhag2024-xa,Mangal2024-rd,Rodrigues2025-iz,Sato2025-ty-rheosindy,Dong2026-cf}. 
Although tensorial descriptions are generally required for predictions under arbitrary flow conditions, standard rheological measurements provide only selected stress components and do not fully determine the state of the extra stress tensor~\cite{Miyamoto2026}.
Nevertheless, such tensorial variables have remained central to constitutive modeling, because they connect constitutive states to measured stress components and provide clear connections to structural descriptors, e.g., conformational orientation and stretch.

Microscopic and mesoscopic theories explain how molecular dynamics contributes to macroscopic stress, often by distinguishing conformational orientation from stretch.
For example, in bead--spring models, the conformation tensor is a convenient state variable in which orientational anisotropy and stretch amplitude are combined. 
The finitely extensible nonlinear elastic model with Peterlin preaveraging (FENE-P) connects molecular stretch to stress relaxation~\cite{Peterlin1966-jg}. 
Tube-based models for entangled polymer chains often distinguish orientation and stretch more explicitly~\cite{Doi1988-cc,Mead1995-nl,Mead1998-ad,Ianniruberto2001-cd,McLeish2002-ro,Marrucci2003-le,Wagner2001-tr,Rolon-Garrido2014-al,Narimissa2019-gr,Yaoita2012-dz}. 
The pom-pom~\cite{McLeish1998-rt}, GLaMM~\cite{Graham2003-ht}, and stretching Rolie--Poly~\cite{Likhtman2003-tr,Renardy2018-yf,Chen2025-to} models retain chain stretch together with orientational dynamics to describe linear and nonlinear stress responses. 
Related approaches have incorporated conformational effects through flow-dependent kinetic coefficients such as the segmental friction coefficient~\cite{Yaoita2012-dz,Ianniruberto2020-jn,Watanabe2021-gp,Sato2021-pt,Miyamoto2023-nw,Psevdos2026-ju}.
In contrast to these stretch--orientation descriptions, constitutive theories for flexible polymers, in particular, often retain stretch either explicitly as an additional state variable or implicitly through the isotropic stress component.
Theories for rigid rods, liquid-crystalline polymers, and suspensions of anisotropic particles are intrinsically based on orientation dynamics without stretch~\cite{Doi1988-cc,Doi1978-aa,Doi1978-rx,Advani1987-cd,de-Gennes1993-xh,Hinch1975-kp,Hinch1976-my,Bird2014-pw,Kamal2024-mb,Palade2026}.

Independently of whether conformational stretch is physically important, this situation raises the question of whether stretch must remain an independently evolving macroscopic state.
In a bead--spring picture, the trace of the conformation tensor represents the conformational stretch contribution, and the trace of the extra stress tensor carries related contributions through the nonlinear spring force.
However, for incompressible fluids, the isotropic part of the extra stress cannot be distinguished from hydrostatic pressure and is not obtained from standard rheological measurements~\cite{Miyamoto2026}.
This motivates a description that retains the influence of stretch without treating either stretch or the stress trace as an additional state variable.

\begin{figure}[tb]
    \centering
    \includegraphics[width=1.0\linewidth]{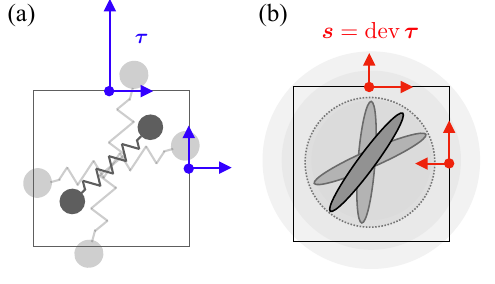}
    \caption{
    Comparison of the extra stress tensor ($\bm\tau$) from the dumbbell system and the deviatoric stress tensor ($\bm s$; $\tr\bm s=0$) from the orientational system with implicit stretch contributions.
    The arrows schematically indicate stress directions; the opposing directions for the deviatoric stress illustrate its traceless character.
    }
    \label{fig:1:motivation}
\end{figure}

In this study, we develop a deviatoric-stress-based constitutive description from the orientational dynamics of a dumbbell model.
We summarize the conventional constitutive description in Sec.~\ref{sec:extra-stress-and-connector}.
In Sec.~\ref{sec:deviatoric-stress-and-orientation}, we separate the stretch and orientational contributions and derive a constitutive equation in which the deviatoric stress is retained as the macroscopic state variable, while the externalized stretch degree of freedom of the conformation enters through an effective stress scale and rotational diffusivity.
Figure~\ref{fig:1:motivation} summarizes the motivation and structure of this reduction.
In Sec.~\ref{sec:effective-coeffs}, we discuss the shear responses of the proposed formulation using the upper-convected Maxwell (UCM) and FENE-P models, examine the physical roles of the effective coefficients, and represent them as functions of stress invariants inferred from steady-shear responses.

\begin{table}[tb]
    \centering
    \caption{
    Correspondence of variables for the dumbbell model in terms of both the connector- and the orientation-descriptions.
    $\bm\tau$ is the extra stress tensor and $\bm s$ is the deviatoric stress tensor.
    $\bm Q$ is the dumbbell connector, $r$ is its length, and $\bm u$ is the orientational vector.
    $\Psi_t(\bm Q)$ is the probability distribution of $\bm Q$.
    $\phi_t(r|\bm u)$ is the radial conditional distribution on $\bm u$, and $\psi_t(\bm u)$ is the orientational distribution.
    $\bm C$ is the conformation tensor and $\bm A$ is the anisotropy tensor.
    $D_t$ and $\sigma_t$ are the radial-distribution dependent coefficients, which are the rotational diffusion coefficient and the stress scale, respectively.
    }
    \begin{tabular}{lll}
    \hline
              & Connector  & Orientation \\ 
    \hline
    Stress    & $\bm\tau=\bm s+z\mathbf{I}/3$ & $\bm s$  \\
    State     & $\bm Q=r\bm u$ & $\bm u$  \\
    Distribution & $\Psi_t(\bm Q)=\phi_t(r|\bm u)\psi_t(\bm u)/r^2$ & $\psi_t(\bm u)$ \\
    Descriptor & $\bm C=\langle\bm Q\bm Q\rangle_{\Psi_t}$ & $\bm A=\langle\bm u\bm u-\mathbf{I}/3\rangle_{\psi_t}$\\
    Stretch    & $\langle r^2\rangle_{\Psi_t}=\tr\bm C$ & Through $D_t$ and $\sigma_t$\\
    \hline
    \end{tabular}
    \label{tab:variables}
\end{table}

\section{Extra stress tensor and dumbbell connector dynamics}
\label{sec:extra-stress-and-connector}

In this section, we review a Maxwell-type constitutive equation and its derivation from the microscopic dynamics of a dumbbell system~\cite{Oldroyd1958-db,Bird1987-la}. 
The three subsections summarize the macroscopic constitutive equation, the microscopic dumbbell connector dynamics, and the resultant conformation-tensor formulation, respectively.

\subsection{Maxwell-type constitutive equation}
\label{sec:total-stress-UCM}
We consider the following Maxwell-type constitutive equation for the extra stress tensor $\bm{\tau}$:
\begin{equation}
    \overset{\triangledown}{\bm\tau}
    +
    \frac{1}{\lambda}\bm\tau
    +
    \bm F(\bm\tau,\bm D)
    =
    2G\bm D.
    \label{eq:total-stress-model}
\end{equation}
The term $\bm{F}$ represents additional contributions, not included in the UCM model.
The upper-convected time derivative is defined as
\begin{equation}
    \overset{\triangledown}{\bm\tau}
    =
    \frac{d{\bm\tau}}{dt}
    -
    \bm\kappa\cdot\bm\tau
    -
    \bm\tau\cdot\bm\kappa^{\mathsf T}.
\end{equation}
For the homogeneous deformations considered here, the material derivative reduces to the ordinary time derivative.
$\lambda$ and $G$ are the relaxation time and modulus, respectively.
$\bm{D}=(\bm{\kappa}+\bm{\kappa}^{\mathrm T})/2$ is the deformation rate associated with the velocity-gradient tensor $\bm{\kappa}$.
Incompressibility $\tr\bm\kappa=\tr\bm D=0$ is assumed.
For an isotropic material, material objectivity restricts $\bm F(\bm{\tau},\bm{D})$ to an isotropic tensor function expressed in terms of invariants and basis tensors constructed from $\bm{\tau}$ and $\bm{D}$~\cite{Lennon2023-xc,Miyamoto2026}.
    
\subsection{Dumbbell dynamics}

Figure~\ref{fig:1:motivation}a illustrates the dumbbell model, and Table~\ref{tab:variables} summarizes the corresponding variables.
The dumbbell configuration is described by the connector vector $\bm Q$ and its probability distribution $\Psi_t(\bm Q)$ at time $t$. 
We can decompose the connector as $\bm Q=r\bm u$ where $r=|\bm Q|$ is the connector length and $\bm u=\bm Q/|\bm Q|$ is the orientation vector.
A convenient macroscopic descriptor of the dumbbell configuration is the following conformation tensor
\begin{equation}
    \bm C = \langle\bm Q\bm Q\rangle_{\Psi_t},
\end{equation}
where $\langle\cdot\rangle_P$ denotes an average with respect to the probability density $P$.
The extra stress tensor is given by the Kramers expression:
\begin{equation}
    \bm \tau 
    = 
    \nu\left\langle h(r)\bm Q\bm Q\right\rangle_{\Psi_t}
    -\nu k_\mathrm{B}T\textbf{I},
    \label{eq:dumbbell-Kramers-stress}
\end{equation}
where $h$ is the spring coefficient, $\nu$ is the number density of dumbbells, $k_{\mathrm B}$ is the Boltzmann constant, $T$ is the temperature, and $\textbf{I}$ is the unit tensor.

The evolution of the conformation tensor follows the configurational probability dynamics of the dumbbell connector.
Following the standard Smoluchowski equation~\cite{Doi1988-cc,Doi2013-lv}, we write the probability current $\bm{J}_{\bm Q}$ in terms of a configuration-space velocity $\dot{\bm Q}$ by the diffusion potential $k_\mathrm{B}T\ln \Psi_t$:
\begin{align}
    \bm J_{\bm Q}
    =
    \dot{\bm Q}\Psi_t
    =
    \left[\bm\kappa\cdot\bm Q
    -\frac{2}{\zeta}\frac{\partial}{\partial\bm Q}(U+k_\mathrm{B}T\ln\Psi_t)\right]\Psi_t,
    \label{eq:dumbbell-connector-evolution}
\end{align}
where $\zeta$ is the friction coefficient of each bead and 
$U=U(r)$ is the spring potential, defined such that $dU/dr=h(r)r$.
The probability density obeys the conservation equation:
\begin{align}
\frac{\partial \Psi_t}{\partial t}
    =
    -\frac{\partial}{\partial\bm Q}\cdot\bm{J}_{\bm Q}
    =
    -\frac{\partial}{\partial\bm Q}\cdot\left(\dot{\bm{Q}}\Psi_t\right).
    \label{eq:dumbbell-continuity}
\end{align}

\subsection{Time evolution of conformation tensor}

From Eqs.~\eqref{eq:dumbbell-Kramers-stress}--\eqref{eq:dumbbell-continuity} and assuming vanishing probability flux at the boundaries of configuration space, we obtain
\begin{equation}
    \overset{\triangledown}{\bm C}=-\frac{4}{\nu\zeta}\bm\tau.
    \label{eq:confomation-stress}
\end{equation}
Equation~\eqref{eq:confomation-stress} relates the upper-convected evolution of the conformation tensor directly to the extra stress.
The derivation is given in the Supplemental Material~\cite{SM}.

For a general nonlinear spring, $\bm{C}$ and $\bm{\tau}$ are not related by a closed constitutive expression.
Specifically, the weighted moment $\langle h(r)\bm{Q}\bm{Q}\rangle_{\Psi_t}$ cannot always be expressed 
solely in terms of $\bm{C}=\langle\bm{Q}\bm{Q}\rangle_{\Psi_t}$.
An additional assumption, such as a preaveraging approximation, is required to obtain a closure for $\bm C$.
In the FENE-P model, the spring coefficient is expressed using the Peterlin preaveraging approximation~\cite{Peterlin1966-jg}:
\begin{equation}
    h=h_\mathrm{eq}f(\tr\bm C);\quad f(\tr\bm C) = \frac{L^2-3}{L^2-\tr\bm C/C_\mathrm{eq}}.
    \label{eq:fene-p-C}
\end{equation}
Here, $C_{\mathrm{eq}}=k_{\mathrm B}T/h_{\mathrm{eq}}$ is the equilibrium value of each diagonal component of the conformation tensor, and $L$ characterizes the maximum extension of a connector.
With this approximation, the stress--conformation relation and conformation-tensor evolution become
\begin{equation}
    \bm\tau = G\left(f\frac{\bm C}{C_\mathrm{eq}}-\textbf{I}\right),\qquad 
    \overset{\triangledown}{\bm C} = - \frac{C_\mathrm{eq}}{\lambda}\frac{\bm\tau}{G},    
    \label{eq:fene-p-conformation-stress}
\end{equation}
where $G=\nu k_{\mathrm B}T$ is the relaxation modulus and $\lambda=\zeta/(4h_{\mathrm{eq}})$ is the dumbbell relaxation time.
Eliminating $\bm{C}$ from Eq.~\eqref{eq:fene-p-conformation-stress} gives the following stress-based equation:
\begin{equation}
    \overset{\triangledown}{\bm\tau}+\left(\frac{f}{\lambda}-\frac{d\ln f}{dt}\right)\bm\tau = 2G\bm D + G\frac{d\ln f}{dt}\textbf{I} .
    \label{eq:FENE-P-total-stress}
\end{equation}

The conventional conformation-tensor description retains the moment $\langle\bm{Q}\bm{Q}\rangle_{\Psi_t}$, in which the stretch factor $r^2$ and the orientation tensor $\bm{u}\bm{u}$ remain combined.
In the following section, we instead integrate the configurational probability flux $\bm{J_Q}$ over the radial coordinate and derive an orientation-based description of the deviatoric stress.

\section{Deviatoric stress tensor and orientational dynamics}
\label{sec:deviatoric-stress-and-orientation}

\begin{figure*}[htbp]
    \centering
    \includegraphics[width=0.90\linewidth]{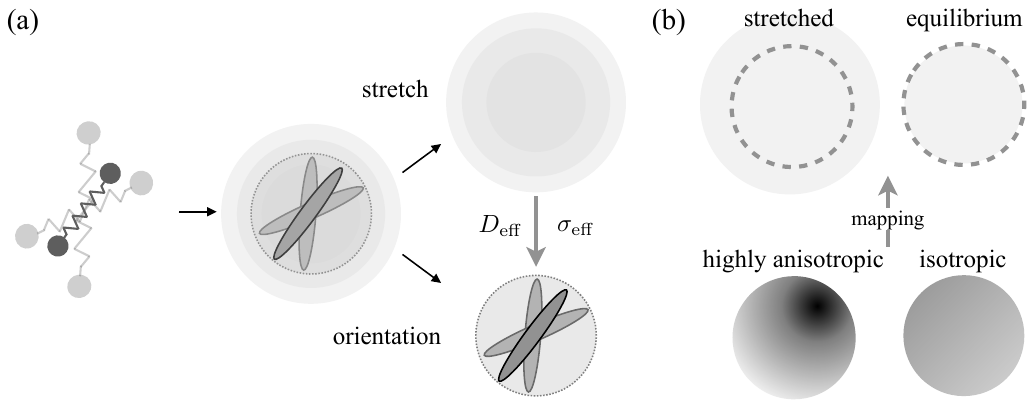}
    \caption{
    Schematic illustration of the orientation-based closure with implicit stretch.
    (a) The dumbbell connector $\bm Q=r\bm u$ is decomposed into radial and orientational degrees of freedom.
    The radial distribution is externalized from the constitutive state, while its effects on the orientational dynamics are retained through an effective rotational diffusivity ($D_\mathrm{eff}$) and stress scale ($\sigma_\mathrm{eff}$).
    (b) Assumed mapping from anisotropy to stretch.
    The stretch-related radial contribution is represented implicitly through anisotropy-dependent effective coefficients rather than through an independently evolving stretch state.
    }
\label{fig:2:theory}
\end{figure*}

In this section, we first decompose the Maxwell-type constitutive equation into its trace and deviatoric parts, and then construct a deviatoric-stress description from the orientational dynamics of the dumbbell model~\cite{Miyamoto2026}.
The constitutive structure is obtained by integrating the dumbbell dynamics over the radial coordinate, and retaining the orientational anisotropy as the structural descriptor.
Unlike the conformation-tensor description, the resultant formulation does not retain radial stretch as an independent state variable.

\subsection{Deviatoric stress constitutive equation}

We decompose the extra-stress tensor $\bm{\tau}$ into its trace and deviatoric parts as
\begin{equation}
    \bm\tau
    =
    \bm s+\frac13z\textbf{I},
    \label{eq:trace-dev-decomposition-of-tau}
\end{equation}
where
\begin{equation}    
    z=\tr\bm\tau,
    \quad
    \bm s=\dev\bm\tau=\bm\tau-\frac{1}{3}(\tr\bm\tau)\textbf{I}.
\end{equation}
Substituting Eq.~\eqref{eq:trace-dev-decomposition-of-tau} into Eq.~\eqref{eq:total-stress-model} gives
\begin{align}
    &\frac{dz}{dt}
    +
    \frac{1}{\lambda}z
    +
    \tr\bm F\left(\bm s+\frac{1}{3}z\textbf{I},\bm D\right)
    =    
    2\bm s:\bm D,
    \label{eq:trace-stress-dyanamics}
    \\
    &\overset{\triangledown_0}{\bm s}
    +
    \frac{1}{\lambda}\bm s
    +
    \dev\bm F\left(\bm s+\frac{1}{3}z\textbf{I},\bm D\right)
    =
    2\left(G+\frac{1}{3}z\right)\bm D.
    \label{eq:deviatoric-stress-dyanamics}
\end{align}
Here, we define the traceless upper-convected derivative of $\bm{s}$ as
\begin{align}
    \overset{\triangledown_0}{\bm s} = \overset{\triangledown}{\bm s}+\frac{2}{3}(\bm s:\bm D)\textbf{I}.
\label{eqn:traceless-ucd}
\end{align}
This definition ensures that the evolution preserves the traceless condition, $\tr\bm{s}=0$.
Equation~\eqref{eq:deviatoric-stress-dyanamics} is an exact consequence of the trace--deviatoric decomposition and shows that the deviatoric dynamics remains coupled to $z$ through both the source term and $\bm{F}$.
In dumbbell models, this trace variable $z$ carries information associated with the stretch of dumbbells through the spring-weighted second moment.

Following our previous study~\cite{Miyamoto2026}, we consider the case where the deviatoric stress tensor is the only macroscopic constitutive state.
 As a minimal closure, we remove the explicit $z$-dependence from Eq.~\eqref{eq:deviatoric-stress-dyanamics} and write
\begin{equation}
    \overset{\triangledown_0}{\bm s}+\frac{1}{\lambda}\bm s + \dev \bm F(\bm s,\bm D)= 2G\bm D.
    \label{eq:deviatoric-stress-closure}
\end{equation}
The advantage of Eq.~\eqref{eq:deviatoric-stress-closure} is that $\bm{s}$ can be reconstructed directly from standard rheometric stress quantities, without estimating $z$. 

We next derive the orientational dynamics obtained from the dumbbell model, and identify the microscopic origin of the terms appearing in the deviatoric-stress equation.

\subsection{Orientational dynamics from the dumbbell model}

Figure~\ref{fig:1:motivation}b illustrates the orientational description, and Table~\ref{tab:variables} summarizes the corresponding variables. 
Following standard treatments of dumbbell dynamics~\cite{Bird1987-la,Doi1988-cc,Advani1987-cd,Phan-Thien2024-su}, we separate the radial and orientational degrees of freedom of a connector vector $\bm{Q}$ by writing $\bm{Q}\bm{Q}=r^2\bm{u}\bm{u}$.
We decompose the configurational probability density into a conditional radial distribution $\phi_t(r|\bm{u})$ and an orientational distribution $\psi_t(\bm{u})$ as
\begin{equation}
    \Psi_t(\bm Q)=\frac{\phi_t(r|\bm u)}{r^2}{\psi_t(\bm u)},
    \label{eq:conditional-prob-decomposition}
\end{equation}
where $r^2$ is the Jacobian factor, with $d\bm Q=r^2drd\bm u$.
To describe the deviatoric stress, we introduce the orientational anisotropy tensor:
\begin{equation}
  \bm A=\dev\langle\bm u\bm u \rangle_{\psi_t}.
\end{equation}
The tensor $\bm{A}$ is the orientational counterpart of the conformation tensor $\bm{C}$, and satisfies $\tr\bm{A}=0$. 
We show below that the conditional radial distribution enters the orientation-based description through two quantities: a stress scale and a rotational diffusion coefficient.

To relate the orientational anisotropy to the deviatoric stress, we take the deviatoric part of the Kramers stress expression in Eq.~\eqref{eq:dumbbell-Kramers-stress}:
\begin{equation}
    \bm s = \nu \dev\left\langle h(r)\bm Q\bm Q\right\rangle_{\Psi_t}=\dev\langle \sigma_t(\bm u)\bm u\bm u \rangle_{\psi_t},
\end{equation}
where
\begin{align}
    \sigma_t(\bm u)=\nu\int h(r)r^2 \phi_t(r|\bm u) dr=\nu\langle h(r) r^2\rangle_{\phi_t}.
    \label{eq:stress-scale-def}
\end{align}
$\sigma_t(\bm{u})$ is an orientation-dependent stress scale, determined by the conditional radial distribution.
It determines how orientational anisotropy is converted into deviatoric stress.

We derive the orientational dynamics from the connector dynamics and probability-conservation equation, with the details given in Supplemental Material~\cite{SM}.
Taking the cross product of Eq.~\eqref{eq:dumbbell-connector-evolution} with $\bm{u}$ gives
\begin{align}
    \bm\omega
    &=
    \bm\Omega-\frac{1}{r^2}\frac{2k_\mathrm{B}T}{\zeta}{\bm{\mathcal{R}}}[\ln\psi_t(\bm u)+\ln \phi_t(r|\bm u)].
    \label{eq:angular-velocity}
\end{align}
Here, $\bm{\omega}=\bm{u}\times \dot{\bm{u}}$ is the angular velocity, $\bm{\Omega}=\bm{u}\times(\bm{\kappa}\cdot\bm{u})$ is the affine contribution, and $\bm{\mathcal{R}}=\bm{u}\times\partial/\partial\bm{u}$ is the rotational operator.
The spring potential does not contribute directly to Eq.~\eqref{eq:angular-velocity} because $U(r)$ depends only on the radial coordinate.
Integrating the configurational probability-conservation represented by Eq.~\eqref{eq:dumbbell-continuity} over the radial coordinate gives the orientational probability-conservation equation
\begin{align}
    \frac{\partial \psi_t(\bm u)}{\partial t}
    =
    -\bm{\mathcal{R}}\cdot\left[\psi_t(\bm u)\int\bm\omega\phi_t(r|\bm u)dr\right]\label{eq:prob-continuity-orientation}.
\end{align}
Substituting Eq.~\eqref{eq:angular-velocity} into Eq.~\eqref{eq:prob-continuity-orientation} gives 
\begin{align}
    \frac{\partial \psi_t}{\partial t}
    =
    -\bm{\mathcal{R}}\cdot(\bm\Omega\psi_t)
    +\bm{\mathcal{R}}\cdot\bm{\mathcal{R}}[D_t(\bm u)\psi_t],
    \label{eq:orientational-prob-dynamics}
\end{align}
where 
\begin{align}
    D_t(\bm u)&=\frac{2k_\mathrm{B}T}{\zeta}\int \frac{1}{r^2} \phi_t(r|\bm u) dr=\frac{2k_\mathrm{B}T}{\zeta}\langle r^{-2}\rangle_{\phi_t}.
    \label{eq:diffusion-coeff}
\end{align}
Here, $D_t(\bm{u})$ is an orientation-dependent rotational diffusion coefficient determined by the radial moment with $r^{-2}$.
This indicates that shorter connectors contribute more strongly to the effective rotational diffusion because they rotate faster than longer, stretched connectors.

The conditional radial distribution enters the orientation-based description through two distinct quantities: the stress scale $\sigma_t(\bm u)$ and the rotational diffusion coefficient $D_t(\bm u)$.
The former determines the conversion from orientational anisotropy to deviatoric stress, whereas the latter governs the rotational diffusion of the orientational distribution.
At this stage, the equations are not yet closed in terms of $\psi_t$ or $\bm A$ alone.

\subsection{Time evolution of orientational anisotropy}

Retaining $\phi_t(r|\bm u)$ would preserve the radial distribution as an additional constitutive state.
As illustrated in Fig.~\ref{fig:2:theory}a, we take an alternative approach: the radial degree of freedom is externalized from the constitutive state, while its effects remain through the stress scale and rotational diffusion.
Figure~\ref{fig:2:theory}b illustrates the assumed relation between the macroscopic orientation and stretch state.
Analogous to Peterlin preaveraging~\cite{Peterlin1966-jg}, we set
\begin{align}
\sigma_t(\bm u) &\approx \sigma_{\rm eff}(\bm A),\label{eq:approx-stress-scale}\\
D_t(\bm u) &\approx D_{\rm eff}(\bm A).\label{eq:approx-diffusion}
\end{align}
Here, $\sigma_{\rm eff}$ and $D_{\rm eff}$ have no explicit dependence on $\bm u$.
Thus, the radial distribution is no longer retained as an explicit constitutive state.
With these approximations, the time evolution equation for $\bm{A}$ becomes
\begin{align}
    \overset{\triangledown_0}{\bm A}
    +6D_\mathrm{eff}\bm A
    +2\dev\langle(\bm u\bm u:\bm D)\bm u\bm u\rangle_{\psi_t}
    =
    \frac23\bm D .
    \label{eq:anisotropy-dynamics-no-closed}
\end{align}
The derivation of Eq.~\eqref{eq:anisotropy-dynamics-no-closed} is given in the Supplemental Material~\cite{SM}.

The affine rotation of the orientation vector generates the moment $\dev\langle(\bm u\bm u:\bm{D})\bm u\bm u\rangle_{\psi_t}$, which couples the orientational anisotropy to the deformation rate.
To close this moment, we retain terms only to first order in the orientational anisotropy $\bm{A}$.
We use the following expansion of the orientational distribution about the isotropic state:
\begin{equation}
\psi_t(\bm u)
=
\psi_{\rm iso}
\left[
1+\frac{15}{2}(\bm u\bm u:\bm A)+\cdots
\right],
\label{eq:weak-anisotropy-expansion}
\end{equation} 
where $\psi_{\mathrm{iso}}=1/(4\pi)$ is the isotropic orientational distribution.
The prefactor $15/2$ ensures that $\dev\langle\bm u\bm u\rangle_{\psi_t}=\bm{A}$ to first order in $\bm{A}$.
Substituting Eq.~\eqref{eq:weak-anisotropy-expansion} into the unclosed orientational moment gives
\begin{equation}
\dev\langle(\bm u\bm u:\bm D)\bm u\bm u\rangle_{\psi_t}
=
\frac{2}{15}\bm D
+
\frac{2}{7}
\operatorname{dev}(\bm A\cdot\bm D+\bm D\cdot\bm A)
+\cdots.
\label{eq:fourth-moment-approx-of-S}
\end{equation}
The isotropic second-, fourth-, and sixth-order moment contractions used in this calculation are given in the Supplemental Material~\cite{SM}.

Under the anisotropy-based approximation in Eq.~\eqref{eq:approx-stress-scale}, the deviatoric stress is related to the orientational anisotropy by
\begin{equation}
    \bm s = \sigma_\mathrm{eff}(\bm A)\bm A.
\end{equation}
$\sigma_{\mathrm{eff}}$ acts as an effective stress scale multiplying the anisotropy $\bm{A}$.
Using Eq.~\eqref{eq:fourth-moment-approx-of-S} in Eq.~\eqref{eq:anisotropy-dynamics-no-closed}, the closed anisotropy equation becomes
\begin{align}
    \overset{\triangledown_0}{\bm A}
    +6D_\mathrm{eff}\bm A
    +\frac{4}{7}\dev(\bm A\cdot\bm D+\bm D\cdot\bm A)=\frac{2}{5}\bm D.
    \label{eq:anisotropy-dynamics-final}
\end{align}
The coefficient $4/7$ originates from the first-order anisotropy expansion~\cite{Advani1987-cd}, and is not expected to remain quantitatively accurate under strongly anisotropic conditions.

Using $\bm{s}=\sigma_{\mathrm{eff}}\bm{A}$, Eq.~\eqref{eq:anisotropy-dynamics-final} can be rewritten in terms of the deviatoric stress as
\begin{align}
\overset{\triangledown_0}{\bm{s}}
+
\frac{1}{\lambda_\mathrm{eff}}\bm{s}
+
\xi\dev(\bm{s}\cdot\bm D+\bm D\cdot\bm{s})
=
2G_\mathrm{eff}\bm D,
\label{eq:reference_S_dynamics}
\end{align}
where the effective coefficients are defined as
\begin{align}
    \lambda_\mathrm{eff} &= \left( 6D_\mathrm{eff} -\frac{d\ln\sigma_\mathrm{eff}}{dt} \right)^{-1},\label{eq:eff-relax-proposed}\\
    G_\mathrm{eff} &= \frac{\sigma_\mathrm{eff}}{5},\\
    \xi &= \frac{4}{7}.
\end{align}
The effective relaxation rate contains two contributions: rotational diffusion through $6D_{\mathrm{eff}}$ and temporal variation of the stress scale through $-d\ln\sigma_{\mathrm{eff}}/dt$.
The effective modulus $G_{\mathrm{eff}}$ represents the conversion from orientational anisotropy to stress magnitude, whereas $\lambda_{\mathrm{eff}}$ represents the resultant relaxation time of the deviatoric stress.
For a steady state, an objective state-dependent representation of the effective coefficients can be constructed from the two invariants of $\bm s$: $I_2=\tr(\bm{s}\cdot\bm{s})$ and $I_3=\tr(\bm{s}\cdot\bm{s}\cdot\bm{s})$.
However, under transient deformation, the term $d\ln\sigma_{\mathrm{eff}}/dt$ may introduce an additional dependence on the deformation rate and deformation history.

In the next section, we use steady-shear material functions of the UCM and FENE-P models to examine the physical roles and possible state dependence of $G_{\mathrm{eff}}$ and $\lambda_{\mathrm{eff}}$.

\section{Modeling of Effective Coefficients}
\label{sec:effective-coeffs}

In this section, we propose a method to estimate the effective coefficients in the deviatoric-stress constitutive equation
from the experimentally accessible rheological data, \emph{i.e.,} the shear stress $\sigma_s$, the normal stress differences $N_1$ and $N_2$.
We first derive the steady-shear material functions of the equation, and examine the UCM and FENE-P models as representative examples.
We then invert the steady-shear relations to infer the effective coefficients as functions of shear rate.
Finally, we construct an illustrative representation of the effective coefficients as functions of deviatoric-stress invariants.
We note that steady shear samples only a one-dimensional trajectory in two-dimensional invariant space, and this invariant-function representation is not unique.

\subsection{Material functions under shear}

We begin with Eq.~\eqref{eq:reference_S_dynamics} obtained in the previous section:
\begin{equation}
\overset{\triangledown_0}{\bm s}
+\frac{1}{\lambda_\mathrm{eff}}\bm s
+\xi\operatorname{dev}(\bm s\cdot\bm D+\bm D\cdot\bm s)
=2G_\mathrm{eff}\bm D.
\notag
\end{equation}
Although the anisotropy closure in the previous section gives $\xi=4/7$, we retain $\xi$ as a general coefficient in the following analysis.

For steady simple shear at a constant shear rate $\dot{\gamma}=\kappa_{xy}$, Eq.~\eqref{eq:reference_S_dynamics} gives the shear stress $\sigma_s=s_{xy}$ and the normal-stress differences $N_1=s_{xx}-s_{yy}$ and $N_2=s_{yy}-s_{zz}$:
\begin{align}
\sigma_\mathrm{s}
&=
G_\mathrm{eff}
\frac{\lambda_\mathrm{eff}\dot\gamma}
{1+K_\xi(\lambda_\mathrm{eff}\dot\gamma)^2},
\label{eq:shear-stress-baseline}
\\
N_1
&=
2G_\mathrm{eff}\frac{(\lambda_\mathrm{eff}\dot\gamma)^2}
{1+K_{\xi}(\lambda_\mathrm{eff}\dot\gamma)^2},
\label{eq:N1-baseline}
\\
N_2
&=
-\frac{\xi}{2}N_1,
\label{eq:N2-baseline}
\end{align}
where $K_\xi={1-(1-\xi)^2}/{3}$.

We should note that, for the typical value $\xi=4/7$, we have $K_\xi=46/49=0.938\ldots\approx1$.
For constant effective coefficients, the resulting steady-shear responses satisfy the approximate correspondences
\begin{align}
 \frac{N_1(\dot\gamma)}{2} \approx G'(\omega)|_{\omega=\dot\gamma} ,\quad \sigma_\mathrm{s}(\dot\gamma) \approx G''(\omega)|_{\omega=\dot\gamma},
\end{align}
where $G'(\omega)$ and $G''(\omega)$ are the storage and loss moduli of the corresponding Maxwell model with the same $G_{\rm eff}$ and $\lambda_{\rm eff}$.
These steady--oscillatory correspondences are closely related to the empirical Cox--Merz and Gleissle--Osaki rules~\cite{Cox1958-vy,Gleissle1980-wb,Osaki1998-bm}.

\subsection{Steady shear responses of the UCM and FENE-P models}
We next apply the corresponding steady-shear representation to the UCM model,
\begin{equation}
    \overset{\triangledown}{\bm\tau}
    +\frac{1}{\lambda}\bm\tau
    = 2G\bm D.
\end{equation}
Taking the trace and deviatoric parts gives
\begin{align}
    &\frac{dz}{dt}+\frac{1}{\lambda}z=2\bm s:\bm D,\\
    &\overset{\triangledown_0}{\bm s}+\frac{1}{\lambda}\bm s=2\left(G+\frac{1}{3}z\right)\bm D.\label{eq:deviatoric-UCM}
\end{align}
Equation~\eqref{eq:deviatoric-UCM} has the form of Eq.~\eqref{eq:reference_S_dynamics} with $\lambda_{\mathrm{eff}}=\lambda$, $G_{\mathrm{eff}}=G+z/3$, and $\xi=0$.
Under steady shear, the trace satisfies $z=2\lambda\dot{\gamma}\sigma_s=N_1$.
Thus, the steady-shear material functions follow from $\lambda_\mathrm{eff}=\lambda$, $G_\mathrm{eff}=G+z/3$, and $\xi=0$, with $z=2\lambda\dot\gamma \sigma_\mathrm{s}=N_1$:
\begin{align}
\sigma_\mathrm{s}
&=
\left(G+\frac{1}{3}N_1\right)\frac{\lambda\dot\gamma}{1+(2/3)(\lambda\dot\gamma)^2}
=G\lambda\dot\gamma,\\
N_1
&=
2\left(G+\frac{1}{3}N_1\right)\frac{(\lambda\dot\gamma)^2}{1+(2/3)(\lambda\dot\gamma)^2}
=2G(\lambda\dot\gamma)^2,\\
N_2&=0.
\end{align}
The increase in the effective modulus $G_{\mathrm{eff}}=G+z/3$ exactly cancels the orientation-derived denominator
$1+(2/3)(\lambda\dot{\gamma})^2$.
Relative to the orientational baseline, the increasing effective stress scale offsets the orientation-induced shear thinning, recovering constant shear viscosity and a quadratic first normal-stress difference.

For the FENE-P model, taking the trace and deviatoric parts of Eq.~\eqref{eq:FENE-P-total-stress} gives
\begin{align}
&
\frac{dz}{dt}
+
\left(\frac{f}{\lambda}-\frac{d\ln f}{dt}\right) z
=
2(\bm s:\bm D) + 3G\frac{d\ln f}{dt},\\
&
\overset{\triangledown_0}{\bm s}
+
\left(\frac{f}{\lambda}-\frac{d\ln f}{dt}\right)\bm s
=
2\left(G+\frac{z}{3}\right)\bm D.
\label{eq:fenep_dev_exact}
\end{align}
Using Eq.~\eqref{eq:fene-p-C}, the FENE factor can be expressed in terms of $z$ and $\bm{s}$ as
\begin{equation}
f=1+\frac{z}{GL^2},
\qquad
\frac{d\ln f}{dt}
=
\frac{2\bm s:\bm D-fz/\lambda}{G(L^2-3)}.
\label{eq:fenep_scalar_feedback}
\end{equation}
Under steady shear, $d\ln f/dt=0$, and Eq.~\eqref{eq:fenep_dev_exact} has the form of Eq.~\eqref{eq:reference_S_dynamics} with $\lambda_{\mathrm{eff}}=\lambda/f$, $G_{\mathrm{eff}}=G+z/3$, and $\xi=0$.
The trace satisfies $z=2\lambda\dot{\gamma}\sigma_s/f=N_1$.
The resultant stresses are $\sigma_\mathrm{s}=G\lambda\dot\gamma/f$, $N_1=2G{(\lambda\dot\gamma)^2}/{f^2}$, and $N_2=0$.
Here, $f\geq1$ is the physical root of $f^3-f^2-2{(\lambda\dot\gamma)^2}/{L^2}=0$.
The UCM and FENE-P models share an increasing effective stress scale, but finite extensibility additionally decreases the effective relaxation time through the factor $f^{-1}$.

Further examples of constitutive equations are provided in Appendix~\ref{sec:app:decomposition-examples}.

\subsection{Effective coefficients inferred from steady shear}

\begin{figure*}[tb]
    \centering
    \includegraphics[width=0.75\linewidth]{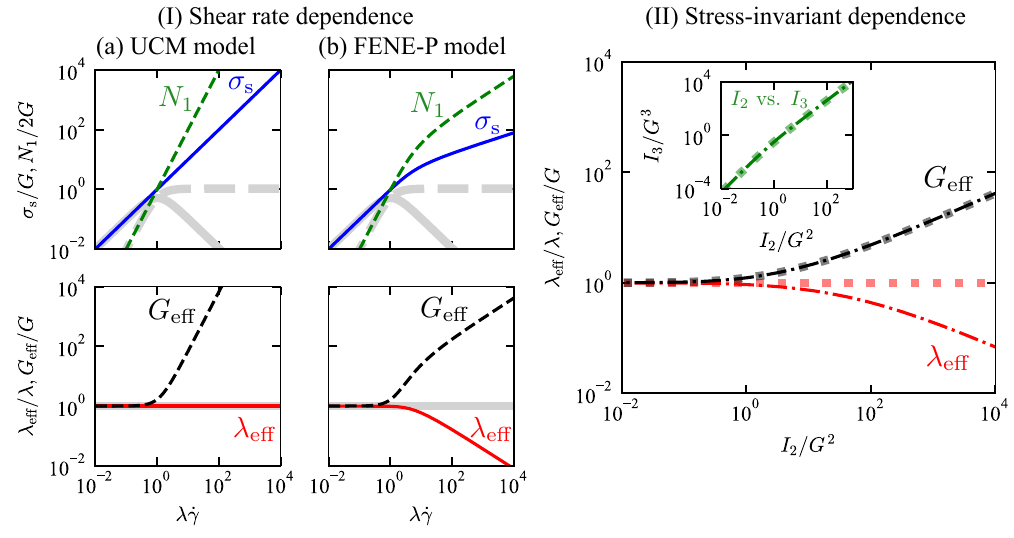}
    \caption{
Steady-shear stress responses and the corresponding effective coefficients.
(I) Shear-rate dependence for (a) the UCM model and (b) the FENE-P model with $L=10$.
The upper panels show the shear stress $\sigma_s$ (blue solid lines) and the first normal-stress difference $N_1$ (green dashed lines), while the lower panels show the effective modulus $G_{\rm eff}$ (black dashed lines) and effective relaxation time $\lambda_{\rm eff}$ (red solid lines).
The thick gray lines denote the orientational baseline with $G_{\rm eff}=G$, $\lambda_{\rm eff}=\lambda$, and $\xi=4/7$.
For the UCM model, $G_{\rm eff}$ increases with shear rate while $\lambda_{\rm eff}$ remains constant; for the FENE-P model, the increase in $G_{\rm eff}$ is suppressed and $\lambda_{\rm eff}$ decreases with shear rate.
(II) Effective coefficients represented as functions of the deviatoric-stress invariant $I_2=\tr(\bm{s}\cdot\bm{s})$.
The dotted and dash-dotted lines show the effective coefficients $G_{\rm eff}$ and $\lambda_{\rm eff}$ for the UCM limit $L\to\infty$ and the FENE-P model with $L=10$, respectively.
The inset shows the relation between $I_2$ and $I_3=\mathrm{tr}(\bm{s}\cdot\bm{s}\cdot\bm{s})$ along the steady-shear trajectories.
    }
    \label{fig:3:understanding}
\end{figure*}

\begin{figure*}[tb]
    \centering
    \includegraphics[width=0.65\textwidth]{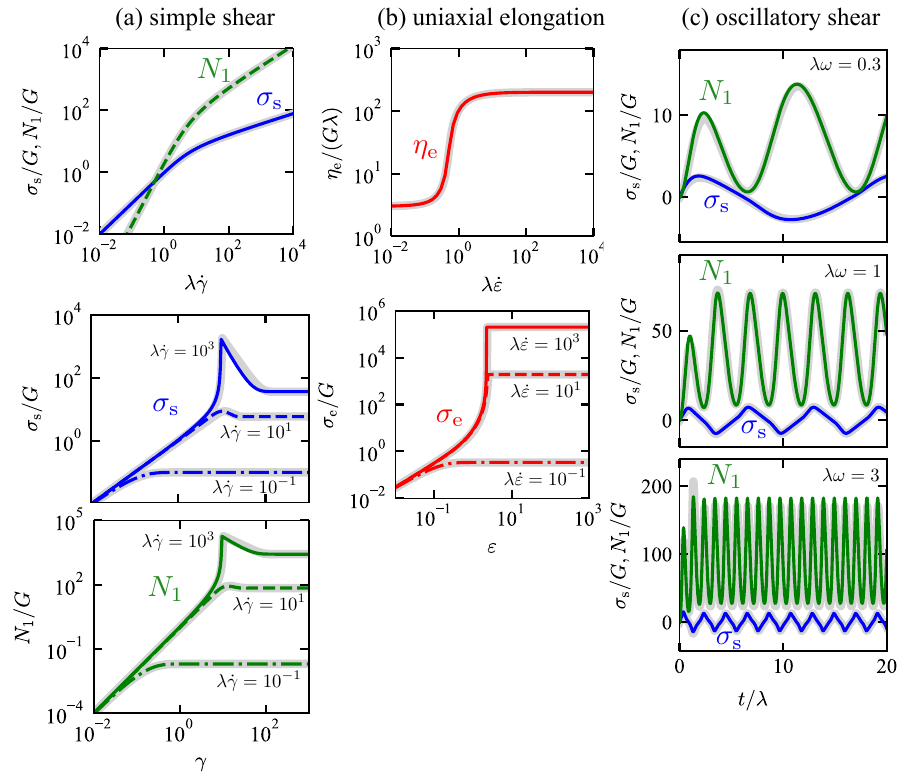}
    \caption{ Reconstruction of stress responses from the original FENE-P model (thick gray lines) and from the deviatoric-stress-based model (thin colored lines) for $L=10$, using Eqs.~\eqref{eq:reference_S_dynamics}, \eqref{eq:lam-eff-from-I2}, and \eqref{eq:G-eff-from-I2}. (a) Simple shear responses at steady state (top) and start-up shear responses of $\sigma_s$ (middle) and $N_1$ (bottom) at $\lambda\dot{\gamma}=10^{-1}$, $10^{1}$, and $10^{3}$. (b) Steady uniaxial elongational viscosity $\eta_\mathrm{e}$ (top) and start-up elongational stress $\sigma_\mathrm{e}$ (bottom) at $\lambda\dot{\epsilon}=10^{-1}$, $10^{1}$, and $10^{3}$. (c) Large-amplitude oscillatory shear responses at $\gamma_0=10$ for $\lambda\omega=0.3$, $1$, and $3$, from top to bottom. Blue, green, and red curves denote $\sigma_\mathrm{s}$, $N_1$, and ($\sigma_\mathrm{e}$, $\eta_\mathrm{e}$), respectively. The steady-shear agreement is imposed by the construction of the effective coefficients, whereas the remaining comparisons test the reconstruction beyond the deformation used for their determination.
    }
    \label{fig:4:prediction}
\end{figure*}

The UCM and FENE-P examples reveal two distinct roles of the effective coefficients in steady shear. 
First, an increase in the effective modulus amplifies stress generation by deformation. 
Second, a decrease in the effective relaxation time accelerates stress relaxation. 
In both models, the increase in the effective modulus also compensates for the orientation-derived denominator.
However, these interpretations are specific to the UCM and FENE-P decompositions, and rheometric observations do not independently determine whether an unobserved trace variable plays the same roles in a real material.
Accordingly, $G_{\mathrm{eff}}$ and $\lambda_{\mathrm{eff}}$ are interpreted as apparent coefficients defined relative to the orientational baseline, rather than as direct measurements of a unique latent stretch variable.

Thus, we define the effective coefficients relative to the orientational baseline directly from steady-shear observables.
Inverting Eqs.~\eqref{eq:shear-stress-baseline}-\eqref{eq:N2-baseline} gives
\begin{align}
    \lambda_\mathrm{eff} &= \frac{1}{\dot\gamma}\frac{N_1}{2\sigma_\mathrm{s}},\label{eq:eff-time-from-shear}\\
    G_\mathrm{eff} &= \frac{2\sigma_\mathrm{s}^2}{N_1}+\frac{K_\xi}{2}{N_1},\label{eq:eff-modulus-from-shear}\\
    \xi &= -2\frac{N_2}{N_1}.
    \label{eq:xi-from-shear}
\end{align}
When all three stress components are available, the coefficients can be evaluated directly from $\sigma_s$, $N_1$, and $N_2$.

Figure~\ref{fig:3:understanding}(I) compares the UCM and FENE-P responses with those of the orientational baseline with constant effective coefficients.
For the orientational baseline, $\sigma_\mathrm{s}$ decreases and $N_1$ saturates at high shear rates.
For the UCM model, $G_{\mathrm{eff}}$ increases with shear rate, whereas $\lambda_{\mathrm{eff}}$ remains constant.
For the FENE-P model, $G_{\mathrm{eff}}$ increases while $\lambda_{\mathrm{eff}}$ decreases with shear rate.
The contrast between the two models distinguishes changes in the stress scale from the finite-extensibility-induced acceleration of stress relaxation.
Both models compensate for the shear thinning predicted by the purely orientational baseline by increasing the effective stress scale, whereas the FENE-P model additionally shortens the effective relaxation time.

The preceding inversion determines the effective coefficients only along the steady-shear trajectory and expresses them as functions of $\dot{\gamma}$.
However, within the state-dependent closure adopted here, the coefficients can be represented as objective functions of the deviatoric-stress state, rather than as explicit functions of shear rate.

\subsection{Invariant-based representation of the effective coefficients}

In the previous subsection, we characterized the effective coefficients along steady-shear responses.
To obtain a constitutive representation independent of the imposed flow protocol, the effective coefficients should be expressed as functions of the current stress state.
Using the invariants of $\bm s$, $I_2=\tr(\bm s\cdot\bm s)$ and $I_3=\tr(\bm s\cdot\bm s\cdot \bm s)$, the effective coefficients can be represented as
\begin{align}
    \lambda_\mathrm{eff}=\lambda_\mathrm{eff}(I_2,I_3),\qquad G_\mathrm{eff}=G_\mathrm{eff}(I_2,I_3),
\end{align}
We use $\xi=0$ in the following UCM and FENE-P reconstruction, because these models produce $N_2=0$ under steady shear (cf. Eq.~\eqref{eq:xi-from-shear}).
Steady-shear data constrain these functions only along a one-dimensional trajectory in the $(I_2,I_3)$ space.
Their continuation away from this trajectory is not uniquely determined by steady-shear data alone.
Under simple shear, the invariants are expressed in terms of the measurable stress components as
\begin{align}
    I_2 = &2\sigma_\mathrm{s}^2+\frac{2}{3}(N_1^2+N_1N_2+N_2^2),\label{eq:I2-shear-stress}\\
    I_3 = &(N_1+2N_2)\sigma_\mathrm{s}^2\nonumber \\
          & +\frac{1}{9}(2N_1^3+3N_1^2N_2-3N_1N_2^2-2N_2^3),
          \label{eq:I3-shear-stress}
\end{align}
These expressions show that both invariants can be evaluated directly from steady-shear rheometric observables without reconstructing the stress trace.

\subsection{Reconstruction of the FENE-P model}

We construct a state-dependent representation of the effective coefficients for the FENE-P model.
Under steady simple shear, substituting the steady-shear stress relations of the FENE-P model ($z=2\sigma_\mathrm{s}^2/G=N_1$ and $N_2=0$) into Eqs.~\eqref{eq:I2-shear-stress} and \eqref{eq:I3-shear-stress} gives the following relations between ($I_2,I_3$) and $z$:
\begin{align}
    I_2 = \frac{2}{3}z^2+Gz,\quad I_3 = \frac{2}{9}z^3+\frac{G}{2}z^2. \label{eq:invariant-trace-FENE-P}
\end{align}
These relations follow from $z=N_1$, $N_2=0$, and the steady-shear stress relations of the FENE-P model.
Steady-shear data do not determine whether $I_2$, $I_3$, or a combination of both provides the appropriate continuation to general stress states. 
For the present reconstruction, we adopt the simpler $I_2$-based representation, and invert the first relation in Eq.~\eqref{eq:invariant-trace-FENE-P}:
\begin{equation}
z_\mathrm{eff}(I_2)
=
\frac{3G}{4}
\left(
\sqrt{1+\frac{8}{3}\frac{I_2}{G^2}}-1
\right).
\label{eq:zeff_definition}
\end{equation}
The effective trace $z_{\mathrm{eff}}(I_2)$ is not an additional constitutive state; it is a function of the retained deviatoric-stress invariant $I_2$.
Substituting $z_{\mathrm{eff}}(I_2)$ into the steady-shear expressions $\lambda_{\mathrm{eff}}=\lambda/f$, $G_{\mathrm{eff}}=G+z/3$, and $f=1+z/(GL^2)$ gives
\begin{align}
    \frac{\lambda_\mathrm{eff}(I_2)}{\lambda}
    &=\left[1+\frac{3}{4L^2}\left(
\sqrt{1+\frac{8}{3}\frac{I_2}{G^2}}-1
\right)\right]^{-1}, \label{eq:lam-eff-from-I2}\\ 
    \frac{G_\mathrm{eff}(I_2)}{G}
    &=1+\frac{1}{4}\left(
\sqrt{1+\frac{8}{3}\frac{I_2}{G^2}}-1
\right).\label{eq:G-eff-from-I2}
\end{align}
The UCM representation is recovered in the limit $L\to\infty$, for which $\lambda_{\mathrm{eff}}=\lambda$ while $G_{\mathrm{eff}}$ retains its $I_2$ dependence.

Figure~\ref{fig:3:understanding}(II) shows the resulting invariant dependence of the effective coefficients.
$G_{\rm eff}$ increases with $I_2$ for both the UCM and FENE-P models, whereas $\lambda_{\rm eff}$ remains constant for the UCM model and decreases for the FENE-P model.
The inset also shows that $I_2$ and $I_3$ are strongly correlated along the steady-shear trajectories, which motivates the present $I_2$-based representation while not uniquely determining its continuation away from steady shear.

Along the steady-shear trajectory used in their construction, these functions exactly reproduce the stress responses of the UCM and FENE-P models.
Figure~\ref{fig:4:prediction} compares the original FENE-P model with the deviatoric-stress reconstruction.
For $L=10$, the reconstructed model is evaluated under steady and transient shear, uniaxial extension, and large-amplitude oscillatory shear.
The model closely reproduces the shear stress and first normal-stress difference under steady and transient shear, as well as the elongational stress under uniaxial extension.
It also reproduces the large-amplitude oscillatory shear responses over the tested frequencies.
Agreement under steady shear is imposed by construction, whereas the transient shear, extension, and oscillatory-shear responses provide tests beyond the data used to determine $z_{\rm eff}(I_2)$.

These results indicate that, for the FENE-P model under the tested deformation conditions, the effects associated with the externalized stretch state can be represented by the state-dependent effective modulus and relaxation time constructed from $z_{\rm eff}(I_2)$.

\section{Discussion}

The present formulation is a closure for dumbbell dynamics rather than an exact transformation. 
After integrating out the radial coordinate, the effects of the radial degree of freedom remain through the effective stress scale $\sigma_{\rm eff}$ and rotational diffusion $D_{\rm eff}$.
Thus, stretch is not neglected, but is externalized from the constitutive state and represented through state-dependent coefficients in the orientational dynamics. 
The traceless convected derivative and the term $\dev(\bm{s}\cdot\bm{D}+\bm{D}\cdot\bm{s})$ originate from affine orientational kinematics, whereas $G_{\mathrm{eff}}$ and $\lambda_{\mathrm{eff}}$ represent the influence of the externalized radial distribution. 
Equation~\eqref{eq:eff-relax-proposed} shows that the effective stress-relaxation rate contains contributions from both rotational diffusion and temporal variation of the stress scale.

In the present reconstruction, $G_{\mathrm{eff}}$ and $\lambda_{\mathrm{eff}}$ were represented as functions of $I_2=\tr(\bm{s}\cdot\bm{s})$. 
This choice was sufficient for the examined FENE-P responses because $I_2$ and $I_3=\tr(\bm{s}\cdot\bm{s}\cdot\bm{s})$ are strongly correlated along the steady-shear trajectory. 
The representation is not unique and steady shear alone cannot determine the coefficient functions over the invariant space. 
The FENE-P equation also contains the coupling invariant $\tr(\bm{s}\cdot\bm{D})$ through $d\ln f/dt$, indicating that explicit deformation-rate dependence may become important under more general transient flows. 
Strong radial--orientational correlations may further require additional deformation invariants or mixed stress--deformation invariants, higher-order tensor terms, memory effects, or multimode extensions.

The main practical advantage of the formulation is that the retained deviatoric stress can be constructed directly from rheometric stress quantities, while the effects of the implicit stretch degree of freedom are represented by effective coefficients. 
This separation provides a constitutive structure suitable for modeling from experimentally accessible stress data.

\section{Conclusion}

We developed a constitutive description of viscoelasticity in terms of the deviatoric stress.
Starting from dumbbell dynamics, we showed that its orientational dynamics is influenced by dumbbell stretch through an effective stress scale and rotational diffusion.
The resultant equation separates common orientational kinematics from explicit stretch dynamics, whose contributions are represented by an effective modulus and relaxation time.
The comparison between the UCM and FENE-P models showed that the implicit stretch contribution can act through two channels: changes in the effective stress scale and in the effective relaxation time.
An invariant-based representation determined from steady shear reproduced the examined transient shear, uniaxial extension, and large-amplitude oscillatory shear responses without evolving an independent trace or stretch state.

The formulation is a closure rather than an exact representation of the full dumbbell dynamics. 
Nevertheless, it demonstrates that stretch effects can be retained implicitly while using the experimentally accessible deviatoric stress as the constitutive state, thereby connecting microscopic interpretation with macroscopic constitutive modeling and data-driven model construction.

\begin{acknowledgments}
This work was supported by the Japan Society for the Promotion of Science (JSPS) KAKENHI under Grant No.~25K23525 awarded to S.M. and 26K01281 to T.S.

OpenAI ChatGPT was used to assist with parts of the mathematical derivations and to polish the English text. 
The authors provided the physical assumptions and mathematical formulation and independently verified all AI-assisted derivations and reviewed the resulting text.
\end{acknowledgments}

\section*{Conflict of Interest}
The authors declare no conflicts of interest.

\section*{Data Availability}
The numerical results can be reproduced from the equations and parameter values provided in the article. 
The data supporting the findings are available from the corresponding author upon reasonable request.

\appendix

\section{Trace--deviatoric decomposition of constitutive models}
\label{sec:app:decomposition-examples}

We provide three additional examples to illustrate how standard constitutive equations modify the effective coefficients, $\lambda_\mathrm{eff}$ and $G_\mathrm{eff}$.

First, for the Phan-Thien--Tanner model~\cite{Thien1977-gs}, the constitutive equation is given by
\begin{equation}
\overset{\triangledown}{\bm\tau}+\frac{1}{\lambda}\exp\left(\alpha_\mathrm{p}\frac{\tr\bm\tau}{G}\right)\bm\tau+\xi_\mathrm{p}(\bm\tau\cdot\bm D+\bm D\cdot\bm\tau)=2G\bm D.
\end{equation}
The corresponding trace and deviatoric equations are
\begin{align}
&\frac{dz}{dt}+ \frac{e^{\alpha_\mathrm{p} z/G}}{\lambda} z = 2(1-\xi_\mathrm{p})\bm s:\bm D \\
&\overset{\triangledown_0}{\bm s}+\frac{e^{\alpha_\mathrm{p} z/G}}{\lambda}\bm s + \xi_\mathrm{p}\dev(\bm s\cdot\bm D+\bm D\cdot\bm s)\notag\\
&=2\left(G+\frac{1-\xi_\mathrm{p}}{3}z\right)\bm D
\end{align}
Thus, the corresponding effective coefficients are $\lambda_\mathrm{eff} = \lambda/e^{\alpha_\mathrm{p} z/G}$, $G_\mathrm{eff} = G+(1-\xi_\mathrm{p})z/3$, and $\xi=\xi_\mathrm{p}$ .
Thus, $\xi_\mathrm{p}$ affects not only the predicted second normal stress differences ($N_2$), but also the effective modulus.

Second, for the Larson model~\cite{Larson1984-hw}, the constitutive equation is given by
\begin{equation}
    \overset{\triangledown}{\bm{\tau}}
    +\frac{1}{\lambda}\bm{\tau}
    +\frac{2\alpha_\mathrm{L}}{3G}
    (\bm{\tau}:\bm{D})
    (\bm{\tau}+G\bm{I})
    =2G\bm{D}.
\end{equation}
The corresponding trace and deviatoric equations are
\begin{align}
    &\frac{dz}{dt}
    +\frac{1}{\lambda}z
    =
    2\left[1-\alpha_\mathrm{L}\left(1+\frac{z}{3G}\right)\right]\bm{s}:\bm{D},
    \\
    &\overset{\triangledown_0}{\bm{s}}
    +\left(\frac{1}{\lambda}+\frac{2\alpha_\mathrm{L}}{3G}\bm{s}:\bm{D}\right)\bm{s}
    =
    2\left(
        G+\frac{z}{3}
    \right)\bm{D}.
\end{align}
The corresponding effective coefficients are $\lambda_\mathrm{eff}=\left({1}/{\lambda}+{2\alpha_\mathrm{L}}\bm{s}:\bm{D}/({3G})\right)^{-1}$, $G_\mathrm{eff}=G+{z}/{3}$, and $\xi=0$.
The $\bm{s}:\bm{D}$ contribution to $\lambda_{\rm eff}$ introduces explicit deformation-rate dependence beyond a stress-only invariant representation.

Third, for the Giesekus model~\cite{Giesekus1982-ov}, the constitutive equation is given by
\begin{equation}
\overset{\triangledown}{\bm\tau}+\frac{1}{\lambda}\bm\tau+\frac{\alpha_\mathrm{g}}{G\lambda}\bm\tau\cdot\bm\tau=2G\bm D.
\end{equation}
The corresponding trace and deviatoric equations are
\begin{align}
&\frac{dz}{dt}+\frac{1}{\lambda}z+\frac{\alpha_\mathrm{g}}{G\lambda}\left(\bm s:\bm s+\frac{z^2}{3}\right) =2\bm s:\bm D,\\
&\overset{\triangledown_0}{\bm s}+\frac{1}{\lambda}\left(1+\frac{2\alpha_\mathrm{g} z}{3G}\right)\bm s+\frac{\alpha_\mathrm{g}}{G\lambda}\dev(\bm s\cdot\bm s)\notag \\
&=2\left(G+\frac{z}{3}\right)\bm D.
\end{align}
Thus, the corresponding effective coefficients are $\lambda_\mathrm{eff} = \lambda/(1+2\alpha_\mathrm{g} z/3G)$, $G_\mathrm{eff} = G+z/3$, and $\xi=0$.
We note that the second normal-stress difference $N_2$ in the Giesekus model is generated by $\dev(\bm{s}\cdot\bm{s})$, rather than by $\xi\dev(\bm{s}\cdot\bm{D}+\bm{D}\cdot\bm{s})$.

\onecolumngrid

\bibliographystyle{apsrev4-2}
\bibliography{main}

@misc{SM,
  note = {See Supplemental Material at [URL will be inserted by publisher].}
}

@BOOK{Doi2013-lv,
  title     = "Soft Matter Physics",
  author    = "Doi, M",
  publisher = "Oxford University Press",
  doi       =  "doi.org/10.1093/acprof:oso/9780199652952.001.0001",
  year      =  2013,
}

@ARTICLE{Bird2014-pw,
  title     = "Dilute rigid dumbbell suspensions in large-amplitude oscillatory
               shear flow: shear stress response",
  author    = "Bird, R B and Giacomin, A J and Schmalzer, A M and Aumnate, C",
  journal   = "J. Chem. Phys.",
  publisher = "AIP Publishing",
  volume    =  140,
  number    =  7,
  pages     =  074904,
  year      =  2014,
  doi       = "10.1063/1.4862899",
}

@misc{Palade2026,
  title         = {Objective kinetic theory for FENE dumbbell suspension},
  author        = {Palade, Liviu Iulian and Giacomin, Alan Jeffrey},
  year          = {2026},
  eprint        = {2607.01157},
  archiveprefix = {arXiv},
  primaryclass  = {physics.flu-dyn},
}

@ARTICLE{Phan-Thien2024-su,
  title     = "Lopsided rigid dumbbell rheology from Langevin equation: A
               graduate tutorial",
  author    = "Phan-Thien, Nhan and Pan, Dingyi and Kanso, Mona A and Giacomin,
               Alan Jeffrey",
  journal   = "Phys. Fluids",
  publisher = "AIP Publishing",
  volume    =  36,
  number    =  9,
  pages     =  091801,
  year      =  2024,
  doi       = "10.1063/5.0220821",
}

@ARTICLE{Larson1984-hw,
  title     = "A constitutive equation for polymer melts based on partially
               extending strand convection",
  author    = "Larson, R G",
  journal   = "J. Rheol.",
  publisher = "Society of Rheology",
  volume    =  28,
  number    =  5,
  pages     = "545--571",
  year      =  1984,
  doi       = "10.1122/1.549761"
}

@ARTICLE{Thien1977-gs,
  title     = "A new constitutive equation derived from network theory",
  author    = "Thien, Nhan Phan and Tanner, Roger I",
  journal   = "J. Nonnewton. Fluid Mech.",
  publisher = "Elsevier BV",
  volume    =  2,
  number    =  4,
  pages     = "353--365",
  year      =  1977,
  doi       = "10.1016/0377-0257(77)80021-9"
}

@ARTICLE{Giesekus1982-ov,
  title     = "A simple constitutive equation for polymer fluids based on the
               concept of deformation-dependent tensorial mobility",
  author    = "Giesekus, H",
  journal   = "J. Nonnewton. Fluid Mech.",
  publisher = "Elsevier BV",
  volume    =  11,
  number    = "1-2",
  pages     = "69--109",
  year      =  1982,
  doi       = "10.1016/0377-0257(82)85016-7"
}

@INPROCEEDINGS{Gleissle1980-wb,
  author    = {Gleissle, Wolfgang},
  title     = {Two simple time--shear rate relations combining viscosity and
               first normal stress coefficient in the linear and non-linear
               flow range},
  booktitle = {Rheology: Volume 2: Fluids},
  editor    = {Astarita, Giovanni and Marrucci, Giuseppe and Nicolais, Luigi},
  pages     = {457--462},
  publisher = {Springer},
  address   = {Boston, MA},
  year      = {1980},
  doi       = {10.1007/978-1-4684-3743-0_85},
  isbn      = {978-1-4684-3743-0},
}

@ARTICLE{Osaki1998-bm,
  title     = "On the estimation of stresses in steady shear flow from the
               dynamic viscoelasticity for polymeric liquids",
  author    = "Osaki, K and Watanabe, H and Inoue, T",
  journal   = "Nihon Reoroji Gakkaishi",
  publisher = "Society of Rheology Japan",
  volume    =  26,
  number    =  1,
  pages     = "49--52",
  year      =  1998,
  doi       = "10.1678/rheology1973.26.1\_49"
}

@ARTICLE{Cox1958-vy,
  title     = "Correlation of dynamic and steady flow viscosities",
  author    = "Cox, W P and Merz, E H",
  journal   = "J. Polym. Sci.",
  publisher = "Wiley",
  volume    =  28,
  number    =  118,
  pages     = "619--622",
  year      =  1958,
  doi       = "10.1002/pol.1958.1202811812"
}

@ARTICLE{Watanabe2021-gp,
  title     = "Revisiting nonlinear flow behavior of rouse chain: Roles of
               {FENE}, friction-reduction, and Brownian force intensity
               variation",
  author    = "Watanabe, Hiroshi and Matsumiya, Yumi and Sato, Takeshi",
  journal   = "Macromolecules",
  publisher = "American Chemical Society (ACS)",
  volume    =  54,
  number    =  8,
  pages     = "3700--3715",
  year      =  2021,
  doi       = "10.1021/acs.macromol.1c00013",
}

@ARTICLE{Ianniruberto2020-jn,
  title     = "Melts of linear polymers in fast flows",
  author    = "Ianniruberto, Giovanni and Marrucci, Giuseppe and Masubuchi,
               Yuichi",
  journal   = "Macromolecules",
  publisher = "American Chemical Society (ACS)",
  volume    =  53,
  number    =  13,
  pages     = "5023--5033",
  year      =  2020,
  doi       = "10.1021/acs.macromol.0c00693",
}

@ARTICLE{Psevdos2026-ju,
  title     = "Flow-induced friction reduction in shear flows of unentangled
               polymer melts",
  author    = "Psevdos, Christos and Ianniruberto, Giovanni",
  journal   = "Macromolecules",
  publisher = "American Chemical Society (ACS)",
  volume    =  59,
  number    =  16,
  pages     = "9154--9161",
  year      =  2026,
  doi       = "10.1021/acs.macromol.6c01365",
}

@ARTICLE{Miyamoto2023-nw,
  title     = "Stretch-orientation-induced reduction of friction in
               well-entangled bidisperse blends: a dual slip-link simulation
               study",
  author    = "Miyamoto, Souta and Sato, Takeshi and Taniguchi, Takashi",
  journal   = "Rheol. Acta",
  publisher = "Springer Science and Business Media LLC",
  volume    =  62,
  number    =  1,
  pages     = "57--70",
  year      =  2023,
  doi       = "10.1007/s00397-022-01378-5",
}

@ARTICLE{Sato2021-pt,
  title     = "A constitutive equation for Rouse model modified for variations
               of spring stiffness, bead friction, and Brownian force intensity
               under flow",
  author    = "Sato, Takeshi and Kwon, Youngdon and Matsumiya, Yumi and
               Watanabe, Hiroshi",
  journal   = "Phys. Fluids",
  publisher = "AIP Publishing",
  volume    =  33,
  number    =  6,
  pages     =  063106,
  year      =  2021,
  doi       = "10.1063/5.0055559",
}

@misc{Miyamoto2026,
  title         = {A deviatoric-stress closure for constitutive modeling of viscoelastic dynamics},
  author        = {Miyamoto, Souta and Moro, Shotaro and Sato, Takeshi and Kato, Shota and Tanabe, Katsuaki and Molina, John J. and Taniguchi, Takashi},
  year          = {2026},
  eprint        = {2609.00635},
  archiveprefix = {arXiv},
  primaryclass  = {cond-mat.soft},
}

@ARTICLE{Kamal2024-mb,
  title     = "Flow and rheology of suspensions of two-dimensional cylindrical
               or anisotropic particles with Navier slip",
  author    = "Kamal, Catherine and Botto, Lorenzo",
  journal   = "Phys. Rev. Fluids",
  publisher = "American Physical Society (APS)",
  volume    =  9,
  number    =  7,
  pages     =  074102,
  year      =  2024,
  doi       = "10.1103/physrevfluids.9.074102",
}

@ARTICLE{Hinch1976-my,
  title     = "Constitutive equations in suspension mechanics. Part 2.
               Approximate forms for a suspension of rigid particles affected by
               Brownian rotations",
  author    = "Hinch, E J and Leal, L G",
  journal   = "J. Fluid Mech.",
  publisher = "Cambridge University Press (CUP)",
  volume    =  76,
  number    =  01,
  pages     =  187,
  year      =  1976,
  doi       = "10.1017/s0022112076003200",
}

@ARTICLE{Hinch1975-kp,
  title     = "Constitutive equations in suspension mechanics. Part 1. General
               formulation",
  author    = "Hinch, E J and Leal, L G",
  journal   = "J. Fluid Mech.",
  publisher = "Cambridge University Press (CUP)",
  volume    =  71,
  number    =  3,
  pages     = "481--495",
  year      =  1975,
  doi       = "10.1017/s0022112075002698",
}

@ARTICLE{Advani1987-cd,
  title     = "The use of tensors to describe and predict fiber orientation in
               short fiber composites",
  author    = "Advani, Suresh G and Tucker, III, Charles L",
  journal   = "J. Rheol.",
  publisher = "Society of Rheology",
  volume    =  31,
  number    =  8,
  pages     = "751--784",
  year      =  1987,
  doi       = "10.1122/1.549945",
}

@ARTICLE{Doi1978-aa,
  title     = "Dynamics of rod-like macromolecules in concentrated solution.
               Part 2",
  author    = "Doi, Masao and Edwards, Sam F",
  journal   = "J. Chem. Soc., Faraday Trans. 2",
  publisher = "Royal Society of Chemistry (RSC)",
  volume    =  74,
  pages     =  918,
  year      =  1978,
  doi       = "10.1039/f29787400918",
}

@ARTICLE{Doi1978-rx,
  title     = "Dynamics of rod-like macromolecules in concentrated solution.
               Part 1",
  author    = "Doi, M and Edwards, S F",
  journal   = "J. Chem. Soc., Faraday Trans. 2",
  publisher = "Royal Society of Chemistry (RSC)",
  volume    =  74,
  pages     =  560,
  year      =  1978,
  doi       = "10.1039/f29787400560",
}

@ARTICLE{McLeish1998-rt,
  title     = "Molecular constitutive equations for a class of branched
               polymers: The pom-pom polymer",
  author    = "McLeish, T C B and Larson, R G",
  journal   = "J. Rheol.",
  publisher = "Society of Rheology",
  volume    =  42,
  number    =  1,
  pages     = "81--110",
  year      =  1998,
  doi       = "10.1122/1.550933"
}

@ARTICLE{Yaoita2012-dz,
  title     = "Primitive chain network simulation of elongational flows of
               entangled linear chains: Stretch/orientation-induced reduction of
               monomeric friction",
  author    = "Yaoita, Takatoshi and Isaki, Takeharu and Masubuchi, Yuichi and
               Watanabe, Hiroshi and Ianniruberto, Giovanni and Marrucci,
               Giuseppe",
  journal   = "Macromolecules",
  publisher = "American Chemical Society (ACS)",
  volume    =  45,
  number    =  6,
  pages     = "2773--2782",
  year      =  2012,
  doi       = "10.1021/ma202525v",
}

@ARTICLE{Wagner2001-tr,
  title     = "The molecular stress function model for polydisperse polymer
               melts with dissipative convective constraint release",
  author    = "Wagner, M H and Rubio, P and Bastian, H",
  journal   = "J. Rheol.",
  publisher = "Society of Rheology",
  volume    =  45,
  number    =  6,
  pages     = "1387--1412",
  year      =  2001,
  doi       = "10.1122/1.1413503",
}

@ARTICLE{Narimissa2019-gr,
  title     = "Review on tube model based constitutive equations for
               polydisperse linear and long-chain branched polymer melts",
  author    = "Narimissa, Esmaeil and Wagner, Manfred H",
  journal   = "J. Rheol.",
  publisher = "Society of Rheology",
  volume    =  63,
  number    =  2,
  pages     = "361--375",
  year      =  2019,
  doi       = "10.1122/1.5064642",
}

@ARTICLE{Rolon-Garrido2014-al,
  title     = "The molecular stress function ({MSF}) model in rheology",
  author    = "Rol\'{o}n-Garrido, V\'{\i}ctor Hugo",
  journal   = "Rheol. Acta",
  publisher = "Springer Science and Business Media LLC",
  volume    =  53,
  number    =  9,
  pages     = "663--700",
  year      =  2014,
  doi       = "10.1007/s00397-014-0787-x",
}

@ARTICLE{Chen2025-to,
  title     = "A critical and constructive analysis of nonstretching
               approximations in a family of Rolie-Poly models",
  author    = "Chen, Vickie and Wang, Brandon and Drucker, Charles T and
               Peterson, Jose",
  journal   = "J. Rheol.",
  publisher = "Society of Rheology",
  volume    =  69,
  number    =  4,
  pages     = "387--407",
  year      =  2025,
  doi       = "10.1122/8.0000941",
}

@ARTICLE{Renardy2018-yf,
  title     = "A singular perturbation study of the Rolie-Poly model",
  author    = "Renardy, Yuriko and Renardy, Michael",
  journal   = "J. Nonnewton. Fluid Mech.",
  publisher = "Elsevier BV",
  volume    =  262,
  pages     = "52--67",
  year      =  2018,
  doi       = "10.1016/j.jnnfm.2018.02.006",
}

@ARTICLE{Likhtman2003-tr,
  title     = "Simple constitutive equation for linear polymer melts derived
               from molecular theory: Rolie--Poly equation",
  author    = "Likhtman, Alexei E and Graham, Richard S",
  journal   = "J. Nonnewton. Fluid Mech.",
  publisher = "Elsevier BV",
  volume    =  114,
  number    =  1,
  pages     = "1--12",
  year      =  2003,
  doi       = "10.1016/s0377-0257(03)00114-9",
}

@ARTICLE{Marrucci2003-le,
  title     = "Flow-induced orientation and stretching of entangled polymers",
  author    = "Marrucci, Giuseppe and Ianniruberto, Giovanni",
  journal   = "Philos. Trans. A Math. Phys. Eng. Sci.",
  publisher = "The Royal Society",
  volume    =  361,
  number    =  1805,
  pages     = "677--87; discussion 687--8",
  year      =  2003,
  doi       = "10.1098/rsta.2002.1156",
}

@ARTICLE{Ianniruberto2001-cd,
  title     = "A simple constitutive equation for entangled polymers with chain
               stretch",
  author    = "Ianniruberto, Giovanni and Marrucci, Giuseppe",
  journal   = "J. Rheol.",
  publisher = "Society of Rheology",
  volume    =  45,
  number    =  6,
  pages     = "1305--1318",
  year      =  2001,
  doi       = "10.1122/1.1402661",
}

@ARTICLE{Mead1998-ad,
  title     = "A molecular theory for fast flows of entangled polymers",
  author    = "Mead, D W and Larson, R G and Doi, M",
  journal   = "Macromolecules",
  publisher = "American Chemical Society (ACS)",
  volume    =  31,
  number    =  22,
  pages     = "7895--7914",
  year      =  1998,
  doi       = "10.1021/ma980127x",
}

@ARTICLE{Mead1995-nl,
  title     = "The reptation model with segmental stretch",
  author    = "Mead, D W and Yavich, D and Leal, L G",
  journal   = "Rheol. Acta",
  publisher = "Springer Science and Business Media LLC",
  volume    =  34,
  number    =  4,
  pages     = "360--383",
  year      =  1995,
  doi       = "10.1007/bf00367152",
}

@ARTICLE{Graham2003-ht,
  title     = "Microscopic theory of linear, entangled polymer chains under
               rapid deformation including chain stretch and convective
               constraint release",
  author    = "Graham, Richard S and Likhtman, Alexei E and McLeish, Tom C B and
               Milner, Scott T",
  journal   = "J. Rheol.",
  publisher = "Society of Rheology",
  volume    =  47,
  number    =  5,
  pages     = "1171--1200",
  year      =  2003,
  doi       = "10.1122/1.1595099"
}

@ARTICLE{Lennon2023-xc,
  title    = "Scientific machine learning for modeling and simulating complex
              fluids",
  author   = "Lennon, Kyle R and McKinley, Gareth H and Swan, James W",
  journal  = "Proc. Natl. Acad. Sci. U. S. A.",
  volume   =  120,
  number   =  27,
  pages    = "e2304669120",
  year     =  2023,
  doi      = "10.1073/pnas.2304669120",
}

@ARTICLE{Sato2025-ty-rheosindy,
  title     = "Rheo-{SINDy}: Finding a constitutive model from rheological data
               for complex fluids using sparse identification for nonlinear
               dynamics",
  author    = "Sato, Takeshi and Miyamoto, Souta and Kato, Shota",
  journal   = "J. Rheol.",
  publisher = "Society of Rheology",
  volume    =  69,
  number    =  1,
  pages     = "15--34",
  year      =  2025,
  doi       = "10.1122/8.0000872",
}

@ARTICLE{Dong2026-cf,
  title     = "A {GENERIC}-guided active learning {SPH} method for viscoelastic
               fluids using Gaussian process regression",
  author    = "Dong, Xuekai and Nieto Simavilla, David and Ouyang, Jie and Wang,
               Xiaodong and Ellero, Marco",
  journal   = "J. Comput. Phys.",
  publisher = "Elsevier BV",
  volume    =  557,
  number    =  114837,
  pages     =  114837,
  year      =  2026,
  doi       = "10.1016/j.jcp.2026.114837",
}

@ARTICLE{Shanbhag2024-xa,
  title     = "Sparse regression for discovery of constitutive models from
               oscillatory shear measurements",
  author    = "Shanbhag, Sachin and Erlebacher, Gordon",
  journal   = "Phys. Fluids",
  publisher = "AIP Publishing",
  volume    =  36,
  number    =  10,
  pages     =  103117,
  year      =  2024,
  doi       = "10.1063/5.0233607",
}

@ARTICLE{Rodrigues2025-iz,
  title     = "Finding the underlying viscoelastic constitutive equation via
               universal differential equations and differentiable physics",
  author    = "Rodrigues, Elias C and Thompson, Roney L and Oliveira, D\'{a}rio
               A B and Ausas, Roberto F",
  journal   = "Eng. Appl. Artif. Intell.",
  publisher = "Elsevier BV",
  volume    =  160,
  number    =  111788,
  pages     =  111788,
  year      =  2025,
  doi       = "10.1016/j.engappai.2025.111788",
}

@ARTICLE{Mangal2024-rd,
  title     = "Data-driven techniques in rheology: Developments, Challenges and
               Perspective",
  author    = "Mangal, Deepak and Jha, Anushka and Dabiri, Donya and Jamali,
               Safa",
  journal   = "Curr. Opin. Colloid Interface Sci.",
  publisher = "Elsevier BV",
  volume    =  75,
  pages     =  101873,
  year      =  2025,
  doi       = "10.1016/j.cocis.2024.101873",
}

@BOOK{Larson1988-nn,
  title     = "Constitutive equations for polymer melts and solutions",
  author    = "Larson, Ronald G",
  publisher = "Butterworth-Heinemann",
  year      =  1988,
  doi       = "10.1016/c2013-0-04284-3",
}

@ARTICLE{Peterlin1966-jg,
  title     = "Hydrodynamics of macromolecules in a velocity field with
               longitudinal gradient",
  author    = "Peterlin, A",
  journal   = "J. Polym. Sci. B",
  publisher = "Wiley",
  volume    =  4,
  number    =  4,
  pages     = "287--291",
  year      =  1966,
  doi       = "10.1002/pol.1966.110040411",
}

@ARTICLE{Oldroyd1958-db,
  title     = "Non-Newtonian effects in steady motion of some idealized
               elastico-viscous liquids",
  author    = "Oldroyd, J G",
  journal   = "Proc. R. Soc. Lond.",
  publisher = "The Royal Society",
  volume    =  245,
  number    =  1241,
  pages     = "278--297",
  year      =  1958,
  doi       = "10.1098/rspa.1958.0083",
}

@BOOK{Brunton2019-jf,
  title     = "Data-driven science and engineering",
  author    = "Brunton, Steven L and Kutz, J Nathan",
  publisher = "Cambridge University Press",
  year      =  2019,
  doi       = "10.1017/9781108380690",
}

@BOOK{de-Gennes1993-xh,
  title     = "The Physics of Liquid Crystals",
  author    = "de Gennes, P G and Prost, J",
  publisher = "Clarendon Press",
  year      =  1993,
}

@BOOK{Doi1988-cc,
  title     = "The theory of polymer dynamics",
  author    = "Doi, M and Edwards, S F",
  publisher = "Clarendon Press",
  year      =  1988,
}

@BOOK{Bird1987-la,
  title     = "Dynamics of polymeric liquids, volume 2: Kinetic Theory",
  author    = "Bird, R B and Curtiss, C F and Armstrong, R C and Hassager, O",
  publisher = "Wiley",
  year      =  1987,
}

@ARTICLE{Bird1995-rx,
  title     = "Constitutive equations for polymeric liquids",
  author    = "Bird, R",
  journal   = "Annu. Rev. Fluid Mech.",
  publisher = "Annual Reviews",
  volume    =  27,
  number    =  1,
  pages     = "169--193",
  year      =  1995,
  doi       = "10.1146/annurev.fluid.27.1.169"
}

@ARTICLE{McLeish2002-ro,
  title     = "Tube theory of entangled polymer dynamics",
  author    = "McLeish, T C B",
  journal   = "Adv. Phys.",
  publisher = "Informa UK Limited",
  volume    =  51,
  number    =  6,
  pages     = "1379--1527",
  year      =  2002,
  doi       = "10.1080/00018730210153216"
}

\end{document}


\title{Supplemental Material for\\
``Polymer viscoelasticity from an orientational closure with implicit stretch''}

\author{Souta Miyamoto}
\affiliation{Department of Chemical Science and Engineering,
Kyoto University, Kyoto 615-8510, Japan}

\author{Takeshi Sato}
\affiliation{Advanced Manufacturing Technology Institute,
Kanazawa University, Kanazawa 920-1192, Japan}

\author{Wanxiang Jin}
\affiliation{Department of Chemical Science and Engineering,
Kyoto University, Kyoto 615-8510, Japan}

\author{Katsuaki Tanabe}
\affiliation{Department of Chemical Science and Engineering,
Kyoto University, Kyoto 615-8510, Japan}
\affiliation{Kyoto MPI Inc., Kyoto 606-8501, Japan}

\author{John J. Molina}
\affiliation{Department of Chemical Science and Engineering,
Kyoto University, Kyoto 615-8510, Japan}

\author{Takashi Taniguchi}
\affiliation{Department of Chemical Science and Engineering,
Kyoto University, Kyoto 615-8510, Japan}

\maketitle

\onecolumngrid

\section{Derivation of Eq.~(7)}
\label{sec:derive-conformation-stress}

Using Eqs.(4)--(6) and assuming vanishing probability flux at the boundaries of configuration space, the conformation-tensor evolution is obtained as follows:
\begin{align}
    \frac{dC_{ij}}{dt}
    &=\int Q_iQ_j \frac{\partial \Psi_t}{\partial t}d\bm Q\\
    &=-\int Q_iQ_j \frac{\partial }{\partial Q_k}\left(\dot{Q}_k\Psi_t \right)d\bm Q\\
    &=\int \frac{\partial Q_iQ_j}{\partial Q_k}\dot{Q}_k\Psi_t d\bm Q\qquad \\
    &=\int(\delta_{ik}Q_j+Q_i\delta_{jk})\left[\kappa_{kl}Q_l-\frac{2}{\zeta}\frac{\partial}{\partial Q_k}(U+k_\mathrm{B}T\ln\Psi_t)\right]\Psi_td\bm Q\\
    &=\int \left[(\kappa_{il}Q_lQ_j+Q_iQ_l\kappa_{jl})\Psi_t-\frac{4h(r)}{\zeta}Q_iQ_j\Psi_t-\frac{2k_\mathrm{B}T}{\zeta}(\delta_{ik}Q_j+Q_i\delta_{jk})\frac{\partial\Psi_t}{\partial Q_k}\right]d\bm Q\\
    &=\kappa_{il}\langle Q_lQ_j\rangle_{\Psi_t}+\langle Q_iQ_l\rangle_{\Psi_t}\kappa_{jl}-\frac{4}{\zeta}\langle h(r)Q_iQ_j\rangle_{\Psi_t}+\frac{2k_\mathrm{B}T}{\zeta}\int \frac{\partial(\delta_{ik}Q_j+Q_i\delta_{jk})}{\partial Q_k} \Psi_td\bm Q\\
    &=\kappa_{il}\langle Q_lQ_j\rangle_{\Psi_t}+\langle Q_iQ_l\rangle_{\Psi_t}\kappa_{jl}-\frac{4}{\zeta}\langle h(r)Q_iQ_j\rangle_{\Psi_t}+\frac{4k_\mathrm{B}T}{\zeta}\delta_{ij}.
\end{align}
Substitution of the Kramers stress expression in Eq.~(4) gives
\begin{align}
    \overset{\triangledown}{\bm C}=\frac{d\bm C}{dt}-\bm\kappa\cdot\bm C-\bm C\cdot\bm\kappa^\mathsf{T}=-\frac{4}{\zeta\nu}[\nu\langle h(r)\bm Q\bm Q\rangle_{\Psi_t}-\nu k_\mathrm{B}T\textbf{I}]
    =-\frac{4}{\zeta\nu}\bm\tau.
\end{align}

\section{Derivation of Eqs.~(21) and (22)}
\label{sec:app-orientational-dynamics}

We derive the orientational dynamics from the dumbbell connector dynamics in Eq.~(5).
From the frictional contribution of Eq.~(5), 
\begin{equation}
    \bm u\times\left(\dot{\bm Q}-\bm\kappa\cdot\bm Q\right)
    =
    \dot{r}\bm u\times\bm u+r\bm u\times\dot{\bm u}-r\bm u\times\bm\kappa\cdot\bm u
    =
    r(\bm\omega-\bm\Omega).
\end{equation}
We can take the rotational operator $\bm{\mathcal{R}}=\bm u\times \partial/\partial{\bm u}$ for the connector vector $\bm Q=r\bm u$:
\begin{equation}
    \frac{\partial}{\partial\bm Q}
    =\bm u\frac{\partial}{\partial r}+\frac{1}{r}(\textbf{I}-\bm u\bm u)\cdot\frac{\partial}{\partial \bm u}
    =\bm u\frac{\partial}{\partial r}-\frac{1}{r}\bm u\times\bm{\mathcal{R}},\qquad 
    \bm u\times \frac{\partial}{\partial\bm Q} =-\frac{1}{r}\bm u\times(\bm u\times\bm{\mathcal{R}}) = \frac{1}{r}\bm{\mathcal{R}}.
\end{equation}
For the potential contribution of Eq.~(5),
\begin{equation}
    \bm u\times\frac{\partial}{\partial\bm Q}[U(r)+k_\mathrm{B}T\ln\Psi_t(\bm Q)]
    =
    \frac{1}{r}\bm{\mathcal{R}}\{U(r)+k_\mathrm{B}T[\ln\phi_t(r|\bm u)+\ln\psi_t(\bm u)-2\ln r]\}
    =
    \frac{k_\mathrm{B}T}{r}\bm{\mathcal{R}}[\ln\phi_t(r|\bm u)+\ln\psi_t(\bm u)].
\end{equation}
Thus, we obtain Eq.~(21).

From Eq.~(6), we calculate 
\begin{align}
    \frac{\partial}{\partial\bm Q}\cdot \left(\dot{\bm Q}\Psi_t\right)
    &= \left(\bm u\frac{\partial}{\partial r}-\frac{1}{r}\bm u\times\bm{\mathcal{R}}\right)\cdot \left(\bm u\dot{r}\Psi_t+r\dot{\bm u}\Psi_t\right)\\
    &=\bm u\frac{\partial}{\partial r}\cdot\left(\bm u\dot{r}\Psi_t\right) -\frac{1}{r}(\bm u\times\bm{\mathcal{R}})\cdot \left(\bm u\dot{r}\Psi_t+r\dot{\bm u}\Psi_t\right)\\
    &=\frac{\partial}{\partial r}\left(\dot{r}\Psi_t\right)+\frac{2}{r}\dot{r}\Psi_t -(\bm u\times\bm{\mathcal{R}})\cdot \left(\dot{\bm u}\Psi_t\right)\\
    &=\frac{1}{r^2}\frac{\partial}{\partial r}\left(r^2\dot{r}\Psi_t\right) + \bm{\mathcal{R}}\cdot(\bm\omega\Psi_t).
\end{align}
We used $\bm{u}\cdot \dot{\bm u}=0$, $d\bm{u}/dt=\bm{\omega}\times\bm{u}$, and $(\bm{u}\times\bm{\mathcal R})\cdot[\bm{u}g(\bm{u})] = -2g(\bm{u})$.
The decomposition into radial and orientational parts is written as
\begin{align}
    \frac{\partial \Psi_t}{\partial t}
    &=
    -\frac{1}{r^2}\frac{\partial}{\partial r}\left(r^2\dot{r}\Psi_t\right)-\bm{\mathcal{R}}\cdot(\bm\omega\Psi_t),\label{eq:prob-continuity-dumbbell-connector-radial-orientation}
\end{align}
Multiplying Eq.~\eqref{eq:prob-continuity-dumbbell-connector-radial-orientation} by $r^2$ and integrating over $r$, the radial-flux term vanishes under the no-flux boundary condition. 
Using
\begin{equation}
\int r^2 \Psi_t dr = \psi_t(\bm u)\int \phi_t(r|\bm u) dr = \psi_t,
\end{equation}
we obtain Eq.~(22).

\section{Derivation of Eq.~(27)}
\label{sec:orientation-tensor-dynamics-derivation}

Equation~(23), together with the approximations in Eqs.~(25) and (26), gives the time evolution of the anisotropy $\bm A$ as follows.
\begin{align}
    \frac{d\bm A}{dt}
    &=\dev\int (\bm u\bm u)\frac{\partial \psi_t}{\partial t}d\bm u \\
    &=-\dev\int (\bm u\bm u)\bm {\mathcal{R}}\cdot\left[\bm \Omega\psi_t-D_\mathrm{eff}\bm {\mathcal{R}}\psi_t\right] d\bm u \\
    &=\dev\int (\bm{\mathcal{R}} \bm u\bm u)\cdot\left[\bm \Omega\psi_t
    -D_\mathrm{eff}\bm{\mathcal{R}}\psi_t\right] d\bm u
\end{align}
For the affine deformation contribution, $\bm\Omega=\bm u\times(\bm\kappa\cdot\bm u)$, 
\begin{align}
    \dev\int ({\mathcal{R}_k} u_i u_j) \Omega_k \psi_t d\bm u
    &=\dev\langle (\epsilon_{ikl}u_ju_l+u_i\epsilon_{jkl}u_l)\epsilon_{kmn}u_m\kappa_{no}u_\alpha\rangle_{\psi_t}\\
    &=\dev\langle (\delta_{lm}\delta_{in}-\delta_{ln}\delta_{mi})u_ju_lu_mu_\alpha\kappa_{n\alpha}+(\delta_{lm}\delta_{jn}-\delta_{ln}\delta_{mj})u_iu_lu_mu_\alpha\kappa_{n\alpha} \rangle_{\psi_t}\\
    &=\dev\langle (\kappa_{i\alpha}u_\alpha u_j + u_i u_\alpha\kappa_{j\alpha}-2u_i u_j u_l u_\alpha\kappa_{l\alpha}) \rangle_{\psi_t}
\end{align}
Thus, we obtain
\begin{align}
    \dev\int (\bm{\mathcal{R}} \bm u\bm u) \bm\Omega \psi_t d\bm u
    &=\bm\kappa\cdot\dev\langle\bm u\bm u\rangle_{\psi_t}+\dev\langle\bm u\bm u\rangle_{\psi_t}\cdot\bm\kappa^{\mathsf T}-2\dev(\langle\bm u\bm u\bm u\bm u\rangle_{\psi_t}:\bm \kappa)\\
    &=\bm\kappa\cdot\bm A+\bm A\cdot\bm\kappa^{\mathsf T}+\frac23\bm D-2\dev(\langle\bm u\bm u\bm u\bm u\rangle_{\psi_t}:\bm D) \\
    &=\bm\kappa\cdot\bm A+\bm A\cdot\bm\kappa^{\mathsf T}+\frac23\bm D-\frac23(\bm A:\bm D)\textbf{I}-2\dev\langle(\bm u\bm u:\bm D)\bm u\bm u\rangle_{\psi_t}.
\end{align}
For the effective diffusion contribution,
\begin{align}
\dev\int (\bm{\mathcal{R}}\bm u\bm u)\cdot (\bm{\mathcal{R}}\psi_t)  d\bm u
=-\dev\int \bm{\mathcal{R}}\cdot(\bm{\mathcal{R}}\bm u\bm u) \psi_t  d\bm u
=-\dev\langle(\bm{\mathcal{R}}\cdot\bm{\mathcal{R}})\bm u\bm u\rangle_{\psi_t}
=6\dev\langle\bm u\bm u\rangle_{\psi_t}=6\bm A.
\end{align}
where the detailed calculation of $(\bm R\cdot\bm R)\bm u\bm u$ is given by
\begin{align}
\mathcal{R}_k\mathcal{R}_k u_i u_j
&=\epsilon_{kmn}u_m\frac{\partial}{\partial u_n}(\epsilon_{ikl}u_ju_l+u_i\epsilon_{jkl}u_l)\\
&=(\delta_{lm}\delta_{in}-\delta_{ln}\delta_{im})u_m\frac{\partial}{\partial u_n}u_ju_l
+(\delta_{lm}\delta{jn}-\delta_{ln}\delta_{jm})u_m\frac{\partial}{\partial u_n}u_iu_l\\
&=u_l\frac{\partial}{\partial u_i}u_ju_l - u_i\frac{\partial}{\partial u_l}u_ju_l
+u_l\frac{\partial}{\partial u_j}u_iu_l - u_j\frac{\partial}{\partial u_l}u_iu_l\\
&=2(\delta_{ij}-3u_iu_j)=-6\left(u_iu_j-\frac{1}{3}\delta_{ij}\right).
\end{align}
Combining these calculations gives Eq.~(27):
\begin{equation}
    \overset{\triangledown_0}{\bm A}
    =\frac{d\bm A}{dt}-\bm\kappa\cdot\bm A-\bm A\cdot\bm\kappa^\mathsf{T}+\frac{2}{3}(\bm A:\bm D)\textbf{I}
    = 
    -2\dev\langle(\bm u\bm u:\bm D)\bm u\bm u\rangle_{\psi_t}+\frac23\bm D 
    -6D_\mathrm{eff}\bm A.
\end{equation}

\section{Derivation of Eq.~(29)}
\label{sec:isotropic-averages}
In three dimensions, the even-order moments of the isotropic orientational distribution $\psi_{\mathrm{iso}}=1/(4\pi)$ are
\begin{align}
    \langle u_i u_j\rangle_{\psi_\mathrm{iso}}&=\frac{1}{3}\delta_{ij},
    \qquad
    \left\langle u_i u_j u_k u_l\right\rangle_{\psi_\mathrm{iso}}
    =
    \frac{1}{15}
    \left(
        \delta_{ij}\delta_{kl}
        +\delta_{ik}\delta_{jl}
        +\delta_{il}\delta_{jk}
    \right),
\end{align}
and
\begin{align}
\left\langle
u_i u_j u_k u_l u_m u_n
\right\rangle_{\psi_\mathrm{iso}}
=
\notag
\frac{1}{105}
(
&\delta_{ij}\delta_{kl}\delta_{mn}
+\delta_{ij}\delta_{km}\delta_{ln}
+\delta_{ij}\delta_{kn}\delta_{lm}
\nonumber\\
+
&\delta_{ik}\delta_{jl}\delta_{mn}
+\delta_{ik}\delta_{jm}\delta_{ln}
+\delta_{ik}\delta_{jn}\delta_{lm}
\nonumber\\
+
&\delta_{il}\delta_{jk}\delta_{mn}
+\delta_{il}\delta_{jm}\delta_{kn}
+\delta_{il}\delta_{jn}\delta_{km}
\nonumber\\
+
&\delta_{im}\delta_{jk}\delta_{ln}
+\delta_{im}\delta_{jl}\delta_{kn}
+\delta_{im}\delta_{jn}\delta_{kl}
\nonumber\\
+
&\delta_{in}\delta_{jk}\delta_{lm}
+\delta_{in}\delta_{jl}\delta_{km}
+\delta_{in}\delta_{jm}\delta_{kl}
).
\end{align}
Using $\tr\bm{D}=0$ and $\tr\bm{A}=0$, the contraction in Eq.~(29) becomes
\begin{align}
    \dev\langle(\bm u\bm u:\bm D)\bm u\bm u\rangle_{\psi_t}
    &=
    \dev\int (\bm u\bm u:\bm D)\bm u\bm u\left(1+\frac{15}{2}\bm u\bm u:\bm A+\cdots\right)\psi_\mathrm{iso}d\bm u\\
    &=
    \dev\left\langle
    (\bm u\bm u:\bm D)\bm u\bm u
    \right\rangle_{\psi_\mathrm{iso}}
    +\frac{15}{2}
    \dev\left\langle
    (\bm u\bm u:\bm D)(\bm u\bm u:\bm A)\bm u\bm u
    \right\rangle_{\psi_\mathrm{iso}}+\cdots\\    
    &=
    \frac{2}{15}\bm D
    +
    \frac{15}{2}\frac{4}{105}\dev(\bm A\cdot\bm D+\bm D\cdot\bm A)+\cdots\\
    &=
    \frac{2}{15}\bm D
    +
    \frac{2}{7}\dev(\bm A\cdot\bm D+\bm D\cdot\bm A)+\cdots.
\end{align}
